\documentclass[aps,prl,twocolumn,superscriptaddress,amsmath,amsfonts,showpacs]{revtex4}
\usepackage{graphicx}
\usepackage{bm} 
\usepackage{amssymb,amsmath}

\begin{document}

%\preprint{}

\title{
Spin nematic state as a candidate of the hidden order phase of URu$_2$Si$_2$
}


\author{Satoshi Fujimoto}
\affiliation{Department of Physics, Kyoto University, Kyoto 606-8502, Japan}


%\thanks{}


\date{\today}

\begin{abstract}
Motivated by the recent discovery of broken four-fold symmetry in the hidden order phase of URu$_2$Si$_2$[R. Okazaki et al., Science {\bf 331}, 439 (2011)],
we examine a scenario of a spin nematic state as a possible candidate of the hidden order phase.
We demonstrate that the scenario naturally explains most of experimental observations, and
furthermore, reproduces successfully the temperature dependence of the spin anisotropy detected by the above-mentioned experiment in a semi-quantitative way. This result provides strong evidence for the realization
of the spin nematic order.
\end{abstract}

\pacs{}


\maketitle

%\section{}
%\subsection{}
%\subsubsection{}

% If in two-column mode, this environment will change to single-column
% format so that long equations can be displayed. Use
%\begin{widetext}
%\end{widetext}

%{\it Introduction ---}
The heavy fermion compound URu$_2$Si$_2$ exhibits a second order phase transition at 
$T_{\rm HO}\approx 17.5$ K. 
In spite of long-standing enormous efforts
in experimental and theoretical studies\cite{PhysRevLett.55.2727,PhysRevLett.56.185,PhysRevLett.68.2680,PhysRevLett.70.2479,PhysRevLett.73.1027,PhysRevLett.81.3723,Chandra,Mineev,PhysRevB.71.054415,PhysRevLett.96.036405,PhysRevLett.103.107202,Kotliar,Harima,bala}, the order parameter of this phase transition has not yet been identified.
The enigmatic features of this so-called "hidden order(HO)" phase are described as follows:
(i) Despite large anomaly in thermodynamic quantities and drastic
reconstruction of the Fermi surfaces at $T=T_{\rm HO}$, there is neither conventional
 magnetic order nor the change of the crystal structure\cite{PhysRevLett.55.2727,PhysRevLett.56.185,PhysRevLett.68.2680,PhysRevB.43.12809,PhysRevLett.87.087203,Takagi,Wiebe,PhysRevLett.98.166404}. (ii) However, under applied pressure, 
 an antiferromagnetic (AF) state with large moment appears, and more surprisingly, the Fermi surfaces in the AF 
 ordered state are almost the same as those found in the HO 
 phase\cite{Amitsuka,dHvA,Elga,PhysRevLett.105.216409,PhysRevB.82.205108}.

Recently, an experimental breakthrough for this issue was achieved by Okazaki et al.\cite{Okazaki}, who found
spontaneous symmetry breaking in the spin space at $T<T_{\rm HO}$.
They reported that the anisotropy of the spin susceptibility in the $xy$-plane,
which is measured by the quantity $\chi_{xy}=\langle S_xS_y\rangle$, becomes nonzero below 
$T_{\rm HO}$. 
%Here $S_{x(y)}$ is the $x$($y$)-component of an electron spin operator.
Since URu$_2$Si$_2$ is tetragonal with four-fold symmetry at $T>T_{\rm HO}$, and the phase transition at $T=T_{\rm HO}$ does not accompany any lattice distortion,
it is reasonable to expect that this symmetry breaking is an 
essential feature of the HO phase, which imposes a crucial constraint on 
possible candidates of the HO parameter.
Motivated by this experimental observation, in this letter, we discuss a possibility that
a spin nematic (SN) state is realized as the hidden order in URu$_2$Si$_2$.
The SN phase is a state with circulating spin currents, but with 
no magnetic moment\cite{Schul,Nerse,com2}.
The circulating spin currents
break spin rotational symmetry, leading to spin anisotropy,
without breaking time-reversal symmetry.
%Thus, in the SN state, anisotropy in the spin space appears dynamically.
We demonstrate that the above-mentioned features of the HO phase are naturally understood within the scenario of the SN order, and furthermore, that the temperature dependence of the spin anisotropy spontaneously generated in the SN state successfully explains the above experimental observation in a semi-quantitative way, providing strong evidence for the realization
of the SN state as the HO phase of URu$_2$Si$_2$.

%{\it Mean field theory of SN order ---}
%We explore basic properties of the SN state in the case of URu$_2$Si$_2$, within
%a mean-field approximation.
We, first, present a mean field analysis for basic properties of the SN state 
applied to the case of URu$_2$Si$_2$.
The SN state is induced by nesting of the Fermi surface as discussed in refs.\cite{Schul,Nerse}. 
In fact, the recent band calculations for URu$_2$Si$_2$ based on an itinerant $f$-electron picture revealed that
there are one electron band 
denoted as $\varepsilon_{\bm{k}_1}$ and one hole band denoted as $\varepsilon_{\bm{k}_2}$\cite{Elga,Biasini}, 
which are nested to each other via the nesting vector $\bm{Q}_0=(0,0,\pi)$:
i.e. $\varepsilon_{\bm{k}+\bm{Q}_01}=-\varepsilon_{\bm{k}2}$
\cite{oppen,oppen2}.
It is noted that $\bm{Q}_0$ is equivalent to
the ordering vector of the large-moment AF state which is observed 
under applied pressure\cite{Amitsuka}.
Moreover, it was pointed out that several experimental results suggest that the energy gap opens on the Fermi surface below $T_{\rm HO}$\cite{Wiebe,yazdani,Schmidt,PhysRevLett.90.067203,Santander}, which is consistent with
the gap generation due to the nesting of the Fermi surface.
Because of these reasons, we employ the scenario that itinerant $f$-electrons undergo
the transition to the SN state triggered by the Fermi surface nesting.
%We denote the electron band and the hole band, respectively, 
%$\varepsilon_{\bm{k}_1}$ and $\varepsilon_{\bm{k}_2}$, which satisfy
%the nesting condition $\varepsilon_{\bm{k}+\bm{Q}_01}=-\varepsilon_{\bm{k}2}$.
%We consider the SN order induced by the nesting of the two Fermi surfaces.
We will discuss the microscopic origin of this instability later.
%$\varepsilon_{\bm{k}_F1}$ and $\varepsilon_{\bm{k}_F2}$
%with the nesting vector $\bm{Q}_0$; i.e. $\varepsilon_{\bm{k}+\bm{Q}_01}=-\varepsilon_{\bm{k}2}$.
%For URu$_2$Si$_2$, we postulate that $\bm{Q}_0=(0,0,\pi)$, and hence $\bm{Q}_0$ is equivalent to
%the ordering vector of the large-moment AF state which occurs under applied pressure.
Since the ordering vectors for the large-moment AF phase and the SN phase are
the same, the feature (ii) mentioned above is naturally understood within our scenario.
%In fact, the SN order can be induced by the Fermi surface nesting
%in the case of URu$_2$Si$_2$, as will be discussed in details later.
The SN phase is a spin-triplet electron-hole pairing state\cite{Schul,Nerse}, and hence 
the order parameter of the SN state for the effective two band model is
\begin{eqnarray}
\mathcal{O}^{\rm SN}_{\sigma\sigma'}(\bm{k})=\langle c^{\dagger}_{\bm{k}1\sigma}c_{\bm{k}+\bm{Q}_02\sigma'}\rangle
=\bm{d}_{12}(\bm{k})
\cdot\bm{\sigma}_{\sigma\sigma'},
\label{SNOP}
\end{eqnarray}
where
$c^{\dagger}_{\bm{k}a\sigma}$ ($c_{\bm{k}a\sigma}$ )
is a creation (an annihilation) operator for an electron in the band $a=1$, $2$ with momentum
$\bm{k}$, spin $\sigma$.
$\bm{d}_{12}(\bm{k})$ is a vector, the direction of which
is parallel to the spin quantization axis of the SN order.  
Because of time-reversal invariance in the SN state, 
we have the condition $\bm{d}^{*}_{12}(-\bm{k})=-\bm{d}_{12}(\bm{k})$.
Furthermore, we impose inversion symmetry, since there is no indication of broken inversion symmetry in
URu$_2$Si$_2$ from experiments. Then, it follows that $\bm{d}^{*}_{12}(\bm{k})=-\bm{d}_{12}(\bm{k})$.
Also, we assume that $\bm{d}_{ab}(\bm{k})$ is symmetric with respect to the exchange of the band 
indices $a$ and $b$. This assumption will be plausibly justified by a microscopic argument given later.
The mean field Hamiltonian for the SN state of the effective two-band model is 
$\mathcal{H}_{\rm MF}=\mathcal{H}_{\rm MF}^{(0)}+\mathcal{H}_{\rm MF}^{(1)}$ with the kinetic energy term
$\mathcal{H}_{\rm MF}^{(0)}=\sum_{\bm{k},\sigma}\sum_{a=1,2}\varepsilon_{\bm{k}a}c^{\dagger}_{\bm{k}a\sigma}c_{\bm{k}a\sigma}$, and the particle-hole pairing term,
\begin{eqnarray}
%&&\mathcal{H}_{\rm MF}=\sum_{\bm{k},\sigma}\sum_{a=1,2}\varepsilon_{\bm{k}a}c^{\dagger}_{\bm{k}a\sigma}c_{\bm{k}a\sigma}
%\nonumber \\
%&&+
\mathcal{H}_{\rm MF}^{(1)}=\sum_{\bm{k},\sigma,\sigma'}\sum_{a,b \atop a\neq b}
[\bm{d}_{ab}(\bm{k})\cdot\bm{\sigma}_{\sigma\sigma'}c^{\dagger}_{\bm{k}+\bm{Q}_0a\sigma}c_{\bm{k}b\sigma'}
+h.c.].
%i\bm{d}_{12}(\bm{k})\cdot\bm{\sigma}_{\sigma\sigma'}
%[c^{\dagger}_{\bm{k}+\bm{Q}_01\sigma}c_{\bm{k}2\sigma'}-c^{\dagger}_{\bm{k}1\sigma'}c_{\bm{k}+\bm{Q}_02\sigma}].
\label{ham}
\end{eqnarray}
%URu$_2$Si$_2$ has the $D_{4h}$ symmetry in the high-temperature disordered state at $T>T_{\rm HO}$.
For URu$_2$Si$_2$, the crystal structure of which has the $D_{4h}$ symmetry, 
the anisotropy of the spin susceptibility  which breaks
 four-fold symmetry down to two-fold symmetry in the $xy$-plane implies that
 the order parameter belongs to two-dimensional (2D)
representation of the $D_{4h}$ symmetry, and in the hidden order phase,
four-fold symmetry in the 2D space is spontaneously broken.
Then, from the symmetry properties of $\bm{d}_{12}(\bm{k})$ discussed above, a possible candidate is
the E$_g$ state with $\bm{d}_{12}(\bm{k})=i(\Delta_1k_yk_z,\Delta_2k_xk_z,0)$ 
(or $i(\Delta_1k_xk_z,\Delta_2k_yk_z,0)$) for small $|\bm{k}|$.
%Possible candidates of the 2D representation are E$_g$ and E$_u$.
%For the E$_g$ state, $\bm{d}(\bm{k})=(\Delta_1k_xk_z,\Delta_2k_yk_z,0)$ for small $|\bm{k}|$, while
%for E$_u$, $\bm{d}(\bm{k})=(\Delta_1k_x,\Delta_2k_y,0)$, where $\Delta_1$ and $\Delta_2$ are determined by the %self-consistent gap equation.
Here, the real parameters $\Delta_1$ and $\Delta_2$ are determined from the self-consistent gap equation.

%In the E$_u$ state, the second term of the Hamiltonian (\ref{ham}) breaks inversion symmetry. Since there is no %indication of broken inversion symmetry from any experimental data, we exclude the possibility of the E$_u$ state, %and focus the following argument on the E$_g$ state.

According to the experiment\cite{Okazaki}, the axis of the Ising-like anisotropy, which is spontaneously generated
in the HO phase, is parallel to the (1,1,0)-direction.
Since our toy model has continuous rotational symmetry in the $xy$-plane, 
the analysis given here is applicable to URu$_2$Si$_2$ by rotating the principle axes by $\pi/4$ around the $z$-axis, i.e. $x'=\frac{1}{\sqrt{2}}(x-y)$ and $y'=\frac{1}{\sqrt{2}}(x+y)$.
Following the experimental observation,
we assume that four-fold symmetry in the $x'y'$-plane is spontaneously broken, resulting in 
the state with $\bm{d}_{12}(\bm{k})\parallel (1,0,0)$ or $(0,1,0)$
in this rotated spin frame. 
%Note that the case of $\bm{d}_{12}(\bm{k})\parallel (0,1,0)$ is also possible.
The direction of $\bm{d}_{12}(\bm{k})$ is determined by
 the detail of the electronic structure and spin-orbit interaction.
However, most of the following results do not depend on it.
%We use the spin basis for which $\sigma_x$ is diagonal,
%$\sigma_x={\rm diag}(1,-1)$.
For the tight-binding model, we choose
\begin{eqnarray}
\bm{d}_{12}(\bm{k})=(i\Delta_1\phi(\bm{k}),0,0), 
\qquad \phi(\bm{k})=\sin k_{\mu'}\sin k_z,
%(i\Delta_1\sin k_{\mu'}\sin k_z,0,0).
\label{snorder}
\end{eqnarray}
%with $\phi(\bm{k})=\sin k_{\mu'}\sin k_z$,
where $\mu'=x'$ or $y'$. 
%Most of the following results hold for both $\mu'=x'$ and $y'$. 

At this stage, we note that the electron-hole pairing term (\ref{ham}) is nonzero only when 
the nesting vector $\bm{Q}_0$ and the momentum dependence of
$\bm{d}_{ab}(\bm{k})$ fulfill the following relation; 
$e^{-i\bm{Q}_0\bm{r}_i}-e^{i\bm{Q}_0\bm{r}_j} \neq 0$ for $\bm{r}_i$, $\bm{r}_j$ satisfying 
$\Delta_{ij}=\sum_{\bm{k}} \phi(\bm{k}) e^{-i\bm{k}(\bm{r}_i-\bm{r}_j)}\neq 0$.
For (\ref{snorder}) and $\bm{Q}_0=(0,0,\pi)$, this is actually fulfilled.
%To see this, we rewrite the pairing term (\ref{ham}) with the state (\ref{snorder}) into the Fourier-transformed form:
%\begin{eqnarray}
%\sum_{i,j}[\Delta_1\Delta_{ij}(\sigma_x)_{\sigma\sigma'} c^{\dagger}_{i1\sigma}c_{j2\sigma'}i(e^{-i\bm{Q}_0\bm{r}_i}-e^{i\bm{Q}_0\bm{r}_j})+h.c.].
%\label{pairterm}
%\end{eqnarray}
%where 
%$\Delta_{ij}=\sum_{\bm{k}} \phi(\bm{k}) e^{-i\bm{k}(\bm{r}_i-\bm{r}_j)}$.
%For the nonzero pairing term Eq.(\ref{pairterm}),
%$i(e^{-i\bm{Q}_0\bm{r}_i}-e^{i\bm{Q}_0\bm{r}_j} )\neq 0$ is required.
%$\Delta_{ij}$ is nonzero only when $\bm{r}_i-\bm{r}_j=(0,\pm a_y,\pm a_z)$
%with $a_{\mu}$ a lattice constant along the $\mu$-axis.
%Thus, for $\bm{Q}_0=(0,0,\pi)$, Eq.(\ref{pairterm}) does not vanish, and hence all the assumptions on
%the order parameter mentioned above are consistent. 
It is instructive to compare this property of the SN phase with the unconventional spin density wave (USDW) state considered in refs. \cite{PhysRevLett.68.2680,PhysRevLett.81.3723}, the order parameter of which is also given by Eq.(\ref{SNOP}) but with $\bm{d}_{12}$ a real even function of $\bm{k}$, because of broken time-reversal symmetry.
It is easy to see that, if $\bm{Q}_0=(0,0,\pi)$, and the momentum dependence of $\bm{d}_{12}(\bm{k})$ is the same as (\ref{snorder}), the particle-hole paring term (\ref{ham}) vanishes,
and hence, this type of the USDW can not be realized.
%For the USDW state, the exponential factor in Eq.(4) is changed to
%$e^{-i\bm{Q}_0\bm{r}_i}+e^{i\bm{Q}_0\bm{r}_j} $. Thus, if $\bm{Q}_0=(0,0,\pi)$, and the momentum dependence of $\bm{d}_{12}(\bm{k})$ is the same as (\ref{snorder}),
%the particle-hole paring term (\ref{ham}) vanishes.
%Thus, this type of the USDW can not be realized.
It is noted that the USDW is more stabilized than the SN phase, if
the order parameter $\bm{d}_{ab}(\bm{k})$
(and hence, the particle-hole pairing interaction) 
is anti-symmetric with respect to the exchange of the band indices $a$, $b$,
or if higher harmonics of the order parameter such as $\phi(\bm{k})=\sin 2k_{\mu'}\sin 2k_z$ is allowed. 
%However, as will be discussed later, it is unlikely that these situations occur.

\begin{figure}[h]
\begin{center}
\includegraphics[width=5.5cm]{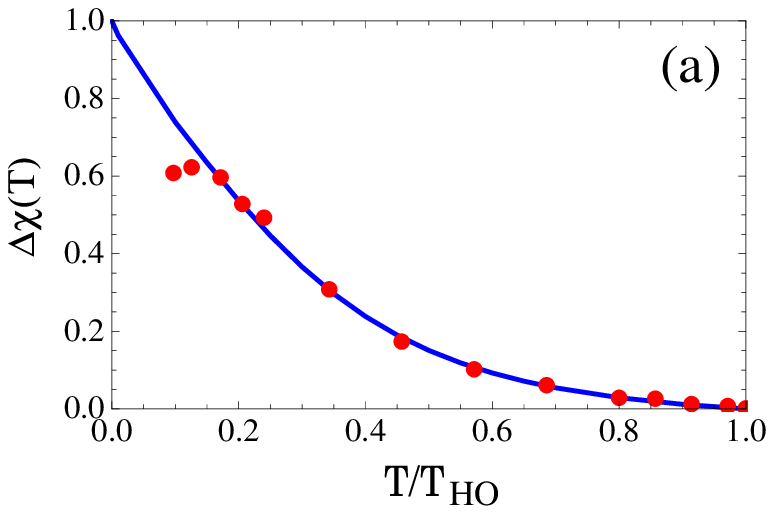}
\includegraphics[width=8cm]{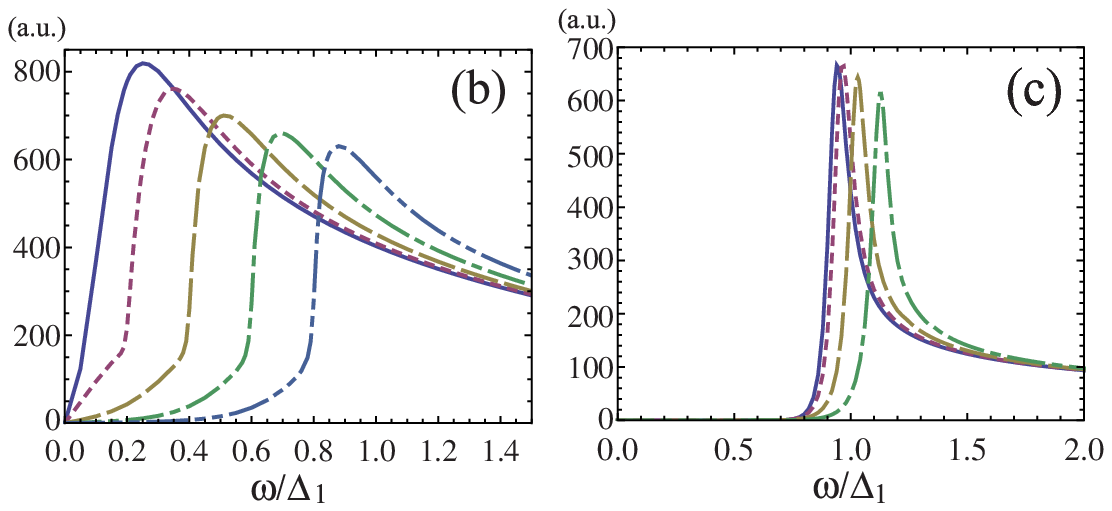}
\caption{ (a) Solid line: anisotropy of the spin susceptibility $\Delta \chi=\chi_{x'x'}-\chi_{y'y'}$ in the SN phase
versus $T/T_{\rm HO}$. The magnitude of $\Delta \chi$ is normalized by
the value at $T=0$. Circle: experimental data quoted from \cite{Okazaki}.
$\Delta_1(0)/T_{\rm HO}=2.6$.
(b) ${\rm Im}\chi^{zz}(\bm{Q}_0,\omega)$ versus $\omega/\Delta_1$. $\mu_{\rm B}H_z/\Delta_1=0$(solid), $0.1$(dotted), $0.2$(dashed), $0.3$(dot-dashed), $0.4$(double-dot-dashed). 
(c) ${\rm Im}\chi^{zz}(\bm{Q}_1,\omega)$ versus $\omega/\Delta_1$. $\mu_{\rm B}H_z/\Delta_1=0$(solid), $0.1$(dotted), $0.2$(dashed), $0.3$(dot-dashed).}
\label{fig:spin-ani}
\end{center}
\end{figure}

We, now, calculate the anisotropy of the uniform spin susceptibility, which characterizes the SN order, and was experimentally
detected in ref.\cite{Okazaki}.
%The SN state breaks the spin rotational symmetry, preserving time reversal symmetry.
The anisotropy in the $x'y'$-plane 
%$\Delta\chi(T)\equiv \chi_{y'y'}(T)-\chi_{x'x'}(T)$ 
for our model is obtained as,
%\begin{eqnarray}
%$\Delta\chi(T)=\mu_{\rm B}^2\sum_{\bm{k}}'\left[\frac{\tanh \frac{E_{\bm{k}}}{2T}}{E_{\bm{k}}}-\frac{1}{2T\cosh^2\frac{E_{\bm{k}}}{2T}}\right]\frac{|\bm{d}_{12}(\bm{k})|^2}{4E^2_{\bm{k}}}$,
%\label{spin-ani}
%\end{eqnarray}
$\Delta\chi(T)\equiv \chi_{y'y'}(T)-\chi_{x'x'}(T)=\mu_{\rm B}^2\sum_{\bm{k}}'\left[\frac{\tanh \frac{E_{\bm{k}}}{2T}}{E_{\bm{k}}}-\frac{1}{2T\cosh^2\frac{E_{\bm{k}}}{2T}}\right]\frac{|\bm{d}_{12}(\bm{k})|^2}{4E^2_{\bm{k}}}
$
where $E_{\bm{k}}=\sqrt{\varepsilon_{\bm{k}1}^2+|\bm{d}_{12}(\bm{k})|^2}$,
and the momentum sum $\sum_{\bm{k}}'$ is taken over the nested part of the Fermi surface satisfying
the condition $\varepsilon_{\bm{k}+\bm{Q}_01}=-\varepsilon_{\bm{k}2}$.
Also, we assume the BCS mean-field-like $T$-dependence of the amplitude of the gap $\Delta_1(T)$.
%Since our effective two-band model is too simple, one cannot  make direct quantitative comparison between
% the theory and the experimental data.
% Nevertheless, it is still possible to compare
% the qualitative behaviors of the temperature dependence of the anisotropy obtained by the theoretical calculation
% with the experimental results, if one normalizes the magnitude of the spin susceptibility properly.
To make a semi-quantitative comparison  with the experimental data without referring to
the details of the band structure,
we consider a ratio $\Delta\chi(T)/\Delta\chi(0)$, for which it is expected that effects of specific band structures approximately cancel out.
There are two parameters in this calculation; 
one is the ratio of the energy gap at $T=0$ to the transition temperature, i.e. 
$\Delta_1(0)/T_{\rm HO}$, and the other is
an overall normalization factor of the magnitude.
According to the recent scanning tunneling microscopy (STM) measurement,
the magnitude of the gap opened on the Fermi surface via the hidden order transition is 
$\Delta_1(0)\sim 4$ meV\cite{yazdani,Schmidt}.
Since $T_{\rm HO}=17.5$ K, we choose the parameter as $\Delta_1(0)/T_{\rm HO}=2.6$.
Then, there is only one fitting parameter, i.e. the overall normalization factor.
We choose the normalization factor to fit 
the theoretical result at $T=6$ K to the experimental data at the same temperature.
%That is, the normalized experimental values of the anisotropy $(\Delta\chi(T)/\Delta\chi(0))_{\rm exp}$ are obtained by
%multiplying the experimental data by a constant factor 
%$\Delta\chi(T=6 {\rm K})/\Delta\chi(0)/(\Delta\chi(T=6 {\rm K}))_{\rm exp}$.
%Here $(...)_{\rm exp}$ is the experimental value.
The calculated result of this one-parameter fitting is shown in FIG.\ref{fig:spin-ani}(a). 
In spite of simplicity of the model, the theoretical result is surprisingly
in good agreement with the experimental observations, at least above $T/T_{\rm HO}\sim 0.15$.
A slight discrepancy at low temperatures may be ascribed to the superconducting transition.
This result provides strong evidence for the realization of the SN state as the HO in URu$_2$Si$_2$.
In FIG.\ref{fig:spin-ani}(a), the $T$-linear behavior of $\Delta \chi(T)$ at low temperatures is seen, which is raised by 
the existence of the line node of the order parameter (\ref{snorder}). 
The experimental data is also consistent with this behavior above $T/T_{\rm HO}\sim 0.15$.

%\begin{figure}[h]
%\begin{center}
%\includegraphics[width=8cm]{Fig-spincor.eps}
%\caption{(a) ${\rm Im}\chi^{zz}(\bm{Q}_0,\omega)$ versus $\omega/\Delta_1$. $\mu_{\rm B}H_z/\Delta_1=0$(solid), $0.1$(dotted), $0.2$(dashed), $0.3$(dot-dashed), $0.4$(double-dot-dashed). 
%(b) ${\rm Im}\chi^{zz}(\bm{Q}_1,\omega)$ versus $\omega/\Delta_1$. $\mu_{\rm B}H_z/\Delta_1=0$(solid), $0.1$(dotted), $0.2$(dashed), $0.3$(dot-dashed).}
%\label{fig:spin-cor}
%\end{center}
%\end{figure}

%%%%added
The above scenario also has important implications for magnetic properties of the HO phase. 
According to neutron scattering measurements, there is a longitudinal spin fluctuation with the wave number
$\bm{q}=\bm{Q}_0$, which exhibits an excitation gap in the HO phase \cite{PhysRevLett.90.067203}.
Remarkably, the magnitude of the spin gap $\sim 1.6$ meV is much smaller than 
the single particle energy gap $\Delta_1\sim 4$ meV observed 
via the STM measurements \cite{yazdani,Schmidt}.
Furthermore, the spin gap increases notably with increasing a magnetic field applied along the $z$-axis, $H_z$; 
e.g. the spin gap for $H_z=17$ T reaches to $2.5$ meV \cite{PhysRevLett.90.067203}.
These properties are well explained by the present model. 
Using the mean field Hamiltonian $\mathcal{H}_{\rm MF}$ and the random phase approximation, 
we calculate the longitudinal spin correlation function 
$\chi^{zz}(\bm{q},\omega)=-i\int^{\infty}_0dt\langle [S^z(\bm{q},t),S^z(-\bm{q},0)]\rangle e^{i\omega t}$ for $\bm{q}=\bm{Q}_0$, which is dominated by the transition between the band $\varepsilon_{\bm{k}1}$ and the band $\varepsilon_{\bm{k}2}$ \cite{com1}. 
%as clarified by the recent band calculations\cite{oppen}.
In FIG. \ref{fig:spin-ani}(b),  the imaginary part of $\chi^{zz}(\bm{Q}_0,\omega)$ is 
plotted as a function of the frequency $\omega/\Delta_1$,
which indicates
the spin excitation gap smaller than the single-electron gap $\Delta_1$.
As the magnetic field $H_z$ increases, the excitation gap increases substantially.
These behaviors are understood as a result of the existence of line nodes of the gap $|\bm{d}_{12}(\bm{k})|$.
The line nodes allow low-energy spin excitations to develop below $\Delta_1$. When the system is in the vicinity of magnetic criticality, a gap structure appears for $\omega<\Delta_1$, which 
is also sensitive to applied magnetic fields,  as shown in FIG. \ref{fig:spin-ani}(b).
These results are in good agreement with the experimental observation obtained in ref.\cite{PhysRevLett.90.067203}.
Conversely, these neutron scattering data strongly imply the existence of line-nodes in the energy gap.
On the other hand, these experiments also revealed that 
in addition to $\bm{Q}_0$,
there is a longitudinal incommensurate spin fluctuations with $\bm{Q}_1=(1.4\pi,0,0)$ \cite{PhysRevB.43.12809},
which may be attributed to the nesting between the hole band $\varepsilon_{\bm{k}2}$ and another hole band denoted as $\varepsilon_{\bm{k}3}$, as suggested from the band calculations \cite{Elga,oppen}.
According to the recent experiments, 
the magnetic excitation for the incommensurate $\bm{Q}_1$
has an energy gap $\sim 4$ meV below $T_{\rm HO}$\cite{PhysRevLett.90.067203,neutron,Wiebe}.
In our scenario, the gap $\sim 4$ meV opens in the hole band $\varepsilon_{\bm{k}2}$
in the SN state, while it does not in the hole band $\varepsilon_{\bm{k}3}$.
Thus, the magnetic excitations due to the transition between these two hole bands should exhibit 
the excitation energy gap $\sim 4$ meV at the wave vector
$\bm{Q}_1$ rather than $2\times 4$ meV.  This explains the above-mentioned experimental result.
To demonstrate this, we calculate the spin correlation function $\chi^{zz}(\bm{Q}_1,\omega)$ 
for the mean field Hamiltonian \cite{com1}.
In this calculation, we postulate that hot spots on the hole band $2$ for this spin fluctuation are located away from the gap-node of $|\bm{d}_{12}(\bm{k})|$; i.e. $|\bm{d}_{12}(\bm{k})|\neq 0$ for $\bm{k}$ satisfying 
$\varepsilon_{\bm{k}+\bm{Q}_1 2}\approx -\varepsilon_{\bm{k}3}$.
Then, the sharp gap edge  appears at $\omega=\Delta_1$ in ${\rm Im}\chi^{zz}(\bm{Q}_1,\omega)$, as shown 
in FIG. \ref{fig:spin-ani}(c). 
In the case with magnetic fields $H_z$,
the gap edge shifts to $\omega\approx \Delta_1+2H_z^2/\Delta_1$ \cite{com1}. 
Thus, the field dependence of ${\rm Im}\chi^{zz}(\bm{Q}_1,\omega)$ is much weaker than
that of ${\rm Im}\chi^{zz}(\bm{Q}_0,\omega)$ (FIG. \ref{fig:spin-ani}(c)).
This is also qualitatively in agreement with the experimental results\cite{PhysRevLett.90.067203,neutron}. 
%%%%%%%%

Some remarks are in order:
%(ii) Our scenario also has an implication for the superconducting state below $T=1.2$ K.
%It was reported that the superconductivity occurs only in the hidden order state, and disappears in the AF state 
%for applied pressure $P>0.7$ GP \cite{Amitsuka}. 
%For the SN state considered above, there is a line node of the gap for the single-electron energy spectrum,
%which allows for low-energy excitations.
%Thus, it is expected that low-energy spin fluctuations with the ordering vector $\bm{Q}_0$ or $\bm{Q}_1$ survive 
%even in the hidden order phase, provided that the nesting vectors connect the portions of the Fermi surface 
%where the line nodes are located.
%In fact, neutron scattering measurements indicate that there still remain spin fluctuations below $T_{\rm HO}$\cite{Wiebe}.
%We deduce that these spin fluctuations mediate the $d$-wave superconducting state proposed in ref.\cite{kasa} 
%in the hidden order phase. In the AF state under applied pressure, the spin fluctuations are well suppressed
%at low temperatures, and thus, the superconductivity does not occur.
(i) In the SN state, there are staggered circulating spin currents \cite{Schul,Nerse}. When a uniform magnetic field parallel to
$(1,1,0)$ or $(1,-1,0)$ in the original frame is applied, and electron spins are polarized, 
the spin currents lead to the staggered orbital currents, which induce staggered moment, as in the
case of the orbital current state \cite{Chandra}. In NMR experiments, broadening of spectra under applied fields
was observed \cite{PhysRevLett.87.087203,Takagi}. 
Its origin may be attributed to the staggered circulating currents of the SN state.
%though it would be also possibly explained by domains of the nematic phase \cite{Okazaki}.
The magnitude of the induced staggered field raised by the circulating currents is estimated as $\sim 1$ Oe for the applied field $H\sim 4$ T, which is consistent with the experimental results\cite{Takagi}.
%though it would be also possibly explained by domains of the nematic phase \cite{Okazaki}.

(ii) Note that the USDW state with the spin quantization axis parallel to $(1,\pm 1,0)$
also exhibits the same spin anisotropy as the SN state, and may explain the experimental data of ref.\cite{Okazaki}.
At this stage, we can not thoroughly exclude the possibility of the USDW as a candidate of the HO, though, as mentioned before, the particle-hole pairing interaction anti-symmetric with respect to the band indices, the origin of which is microscopically unclear, is required for its realization.

(iii) The SN order parameter couples to a magnetic field $\bm{H}$ as $\sim (\bm{d}_{12}\cdot \bm{H})^2$ in the free energy.  This implies that the nonlinear susceptibility $\chi_3$ for a magnetic field $\bm{H}$ satisfying $\bm{d}_{12}\cdot \bm{H}\neq 0$
exhibits a discontinuous jump at $T=T_{\rm HO}$ for the SN order (\ref{snorder}).
However, such an anomaly was experimentally observed not for the in-plane field, but for $\bm{H} \parallel  z$-axis \cite{PhysRevLett.68.2680}. 
The discontinuous jump for $\bm{H} \parallel z$-axis may be explained by postulating that
the SN vector $\bm{d}_{12}$ is not confined in the $xy$ plane, 
%as assumed in Eq.(\ref{snorder}), 
but rather has a nonzero out-of-plane component. 
This scenario may also explain the change of the slope of the linear spin susceptibility at $T=T_{\rm HO}$ for $\bm{H} \parallel z$-axis.
However, the absence of the jump for the in-plane field in the experiment remains to be resolved.
It is desirable to re-examine the measurement of the nonlinear susceptibility using recent high-quality samples.

(iv) According to specific heat measurements in magnetic fields parallel to the $z$-axis $H_z$,
the specific heat jump $\Delta C$ at $T=T_{\rm HO}$ is almost
constant as a function of $H_z$, though the transition temperature decreases
as $T_{\rm HO}(H_z)-T_{\rm HO}(0)\propto -H_z^2$ \cite{dijk,jaime}.
%with $a_0$ a constant \cite{dijk,jaime}.
These behaviors are explained by the SN scenario. 
The decrease of $T_{\rm HO}$ is understood as a result of the decrease of the particle-hole pairing interaction
due to the magnetic field \cite{com1}. 
Also, the absence of the change of $\Delta C$ under applied fields is explained by taking into account
the selfenergy correction arising from interactions with AF spin fluctuations \cite{com1}. 
Within this scenario, the specific heat jump is given by $\Delta C\sim \xi^2 T_{\rm HO}$, where
$\xi$ is the correlation length for AF spin fluctuations.
If the system is in the vicinity of the AF critical point, i.e. $\xi^2\sim 1/T$,
$\Delta C$ is constant, though $T_{\rm HO}$ is decreased by the magnetic field.
  
(v) Experimental studies reported that under applied pressure, the discontinuous phase transition from the HO state to the large moment AF state occurs \cite{Amitsuka}, while the Fermi surfaces in both phases are almost the same \cite{dHvA,PhysRevLett.105.216409,PhysRevB.82.205108,PhysRevLett.98.166404}. According to our scenario, there are competing AF spin fluctuations and fluctuations toward the SN phase transition in the system. (see below and Supplemental Material for details) Thus, applied pressure induces the change of effective couplings between electrons and these fluctuations, resulting in the transition
to the AF phase. Since the order parameters of these phases have the different symmetries that are not compatible with continuous phase transition, the type of the phase transition is first-order, which is in agreement 
with experiments \cite{Amitsuka,Takagi}. 
We stress that the reconstructed Fermi surfaces in the SN phase and the AF phase are the same in our scenario, 
because both of them are reconstructed by the same nesting vector $\bm{Q}_0$.  This is also
in agreement with experimental observations \cite{dHvA,Elga,PhysRevLett.105.216409,PhysRevB.82.205108,PhysRevLett.98.166404}.
  
%{\it Microscopic mechanism of the SN order---}
Finally, we discuss microscopic mechanism of the SN order for URu$_2$Si$_2$ \cite{com1}.
%We examine a scenario that the SN order is induced by the Fermi surface nesting with the nesting vector $\bm{Q}_0$ and electron correlation effects in a multi-band system.
We consider a scenario that orbital fluctuations 
associated with the Fermi surface nesting yields the SN order.
%postulate that the incommensurate AF spin fluctuation due to the nesting vector $\bm{Q}_1$ gives rise to an effective
%interaction which mediates the particle-hole pairing (\ref{SNOP}) with the $d_{yz}$ (or $d_{xz}$) symmetry, Eq.(\ref{snorder}).
%The minimal model consists of one electron band  $\varepsilon_{\bm{k}1}$ and two hole bands $\varepsilon_{\bm{k}2}$ and $\varepsilon_{\bm{k}3}$.
%We also assume that electron-electron interactions between anti-parallel spin states dominate over
%those between parallel spin states, and hence, neglect the latter contributions.
The minimal model consists of the three bands $\varepsilon_{\bm{k}a}$ with $a=1,2,3$, and mutual Coulomb interaction
between electrons. 
%electron system with the Hamiltonian, 
%$\mathcal{H}=\mathcal{H}_{\rm kin}+\mathcal{H}_{\rm int}$ with
%$\mathcal{H}_{\rm kin}=\sum_{\bm{k},\sigma}\sum_{a=1,2,3}\varepsilon_{\bm{k}a}c^{\dagger}_{\bm{k}a\sigma}c_{\bm{k}a\sigma}$
%and,
%%%%%%%%%%%%%%%%%%%%%%%%%%%%
%$\mathcal{H}=\sum_{\bm{k},\sigma}\sum_{a=1,2,3}\varepsilon_{\bm{k}a}c^{\dagger}_{\bm{k}a\sigma}c_{\bm{k}a\sigma}+
%\sum_{\bm{k},\bm{k}',\bm{q} \atop a,b,c,d,\sigma_i}U^{abcd}_{\sigma_1\sigma_2\sigma_3\sigma_4}
%c^{\dagger}_{\bm{k}+\bm{q}a\sigma_1}c^{\dagger}_{\bm{k}'-\bm{q}b\sigma_2}
%c_{\bm{k}'c\sigma_3}c_{\bm{k}d\sigma_4}$.
%%%%%%%%%%%%%%%%%%%%%%%%%%%%
%$\mathcal{H}=\sum_{\bm{k},\sigma}\sum_{a=1,2,3}\varepsilon_{\bm{k}a}c^{\dagger}_{\bm{k}a\sigma}c_{\bm{k}a\sigma}+\sum_{\bm{k},\bm{k}',\bm{q}}\sum_{a,b,c,d}U^{abcd}c^{\dagger}_{\bm{k}+\bm{q}a\uparrow}c^{\dagger}_{\bm{k}'-\bm{q}b\downarrow}c_{\bm{k}'c\downarrow}c_{\bm{k}d\uparrow}$.
%%%%%%%%%%%%%%%%%%%%%%%%%%%%
%\begin{eqnarray}
%&&\mathcal{H}=\sum_{k,\sigma}\sum_{a=1,2,3}\varepsilon_{\bm{k}a}c^{\dagger}_{\bm{k}a\sigma}c_{\bm{k}a\sigma}
%\nonumber \\
%&& +
%\mathcal{H}_{\rm int}=\sum_{\bm{k},\bm{k}',\bm{q}}\sum_{a,b,c,d}U^{abcd}c^{\dagger}_{\bm{k}+\bm{q}a\uparrow}c^{\dagger}_{\bm{k}'-\bm{q}b\downarrow}
%c_{\bm{k}'c\downarrow}c_{\bm{k}d\uparrow}.
%\end{eqnarray}
The SN order considered here arises from the particle-hole pairings in the $d$-wave channel 
which are formed between between the electron band 1 and the hole band 2. 
The Fermi surface nesting with the nesting vector $\bm{Q}_0$ between these bands leads to this instability.
%For simplicity, we assume perfect nesting between the band 1 and the band 2.
%Then, the self-consistent gap equation for the SN state has the BCS form,
%$\Delta_1=V_d\sum_{\bm{k}}\phi^2(\bm{k})\frac{\tanh(\frac{E_{{\bm k}}}{2T})}{E_{{\bm k}}}\Delta_{1}/\sum_{\bm{k}}\phi^2(\bm{k})$, where $V_d$ is an effective pairing interaction in the $d_{\mu'z}$-wave channel.
%$V_d$ 
An effective pairing interaction in the $d_{\mu'z}$-wave channel
is mediated via orbital fluctuations arising from the Fermi surface nesting between 
$\varepsilon_{\bm{k}3(1)}$ and $\varepsilon_{\bm{k}2}$
with the nesting vector $\bm{Q}_1$ ($\bm{Q}_0$).
If the orbital fluctuations are sufficiently strong, the SN order is stabilized \cite{com1}.
 
In conclusion, we have demonstrated that the SN state induced by the Fermi surface nesting is a promising candidate of the HO phase of URu$_2$Si$_2$, since it successfully explains most of experimental observations including
the recent experiment on four-fold symmetry-breaking.
%the four-fold symmetry-breaking in the HO phase detected by the recent experiment, and furthermore, resolves most of puzzles of the HO.
 
The author thanks Y. Matsuda and T. Shibauchi for invaluable discussions, and 
kindly providing the author their experimental data.
He is also indebted to A. V. Balatsky, P. M. Oppeneer, and A. Yazdani for useful discussions.
This work is supported by the Grant-in-Aids for
Scientific Research from MEXT of Japan
(Grants No.19052003 and No.21102510).

\bibliography{spin-nematic}

%\begin{thebibliography}{99}
%\bibitem{palstra} T. T. M. Palstra, A. A. Menovsky, J. van den Berg, A. J. Dirkmaat, P. H. Kes, G. J. Nieuwenhuys, and J. A. Mydosh, Phys. Rev. Lett. {\bf 55}, 2727 (1985).

%\bibitem{maple} M. B. Maple

%\bibitem{broholm} C. Broholm, J. K. Kjems, W. J. L. Buyers, P. Matthews, T. T. M. Palstra, A. A. Menovsky, and J. A. Mydosh,  Phys. Rev. Lett. {\bf 58}, 1467 (1987).

%\bibitem{okazaki} R. Okazaki, T. Shibauchi, H. J. Shi, Y. Haga, T. D. Matsuda, E. Yamamoto, Y. Onuki, H. Ikeda, and Y. Matsuda, .... 

%\bibitem{theory1} V. Barzykin and L. P. Gorkov, 

%\end{thebibliography}

\section{Supplemental Material}
\subsection{A. Spin correlation functions in the spin nematic state}

In this section, we present the calculations of the dynamical longitudinal spin correlation functions 
$\chi^{zz}(\bm{q},\omega)$ in the SN state, particularly focusing on the modes with $\bm{q}=\bm{Q}_0$ or $\bm{Q}_1$, for which spin fluctuations are prominent.
The minimal model consists of the three bands, $\varepsilon_{\bm{k}1}$, $\varepsilon_{\bm{k}2}$, and $\varepsilon_{\bm{k}3}$, with the particle-hole pairing term, and 
the mutual Coulomb interaction between electrons in these bands. 
The Hamiltonian is 
\begin{eqnarray}
\mathcal{H}&=&\sum_{\bm{k},\sigma,a}\varepsilon_{\bm{k}a}c^{\dagger}_{\bm{k}a\sigma}c_{\bm{k}a\sigma}
\nonumber \\
&+&\sum_{\bm{k},\sigma,\sigma'}\sum_{a,b \atop a\neq b}
[\bm{d}_{ab}(\bm{k})\cdot\bm{\sigma}_{\sigma\sigma'}c^{\dagger}_{\bm{k}+\bm{Q}_0a\sigma}c_{\bm{k}b\sigma'}
+h.c.] \nonumber \\
&+&\sum_{\bm{k},\bm{k}',\bm{q} \atop a,b,c,d,\sigma_i}U^{abcd}_{\sigma_1\sigma_2\sigma_3\sigma_4}
c^{\dagger}_{\bm{k}+\bm{q}a\sigma_1}c^{\dagger}_{\bm{k}'-\bm{q}b\sigma_2}
c_{\bm{k}'c\sigma_3}c_{\bm{k}d\sigma_4},
\label{eq:aham}
\end{eqnarray}
with $a,b,c,d=1,2,3$. The SN order occurs for electrons in the band 1 and the band 2, i.e. 
$|\bm{d}_{12}(\bm{k})|=|\bm{d}_{21}(\bm{k})|\neq 0$, and $|\bm{d}_{ab}(\bm{k})|=0$ for other $a$, $b$.

Following the results of the recent band calculation\cite{oppen2},
we postulate that the three bands, $\varepsilon_{\bm{k}1}$, $\varepsilon_{\bm{k}2}$, and $\varepsilon_{\bm{k}3}$,
have the orbital character with $J_z=\pm 5/2$, $\pm 3/2$, and $\pm 1/2$, respectively,
and that each of the single electron states in the $a$-th band with spin $\sigma$ denoted as $|a \ \sigma \rangle$ 
approximately corresponds to
the orbital state $|J_z\rangle$. That is, 
\begin{eqnarray}
&&|1 \uparrow \rangle=|J_z=5/2\rangle, \qquad |1 \downarrow \rangle=|-5/2\rangle,  \label{orb1}\\
&&|2 \uparrow \rangle=|3/2\rangle, \qquad |2 \downarrow \rangle=|-3/2\rangle, \label{orb2}\\
&&|3 \uparrow \rangle=|1/2\rangle, \qquad |3 \downarrow \rangle=|-1/2\rangle.\label{orb3}
\end{eqnarray}
The interaction term of (\ref{eq:aham}) conserve the total $J_z$ in each scattering process.
We, furthermore, assume that 
in the interaction term of (\ref{eq:aham}), interactions between electrons with anti-parallel spins
dominates over those between electrons with parallel spins, and neglect the latter.
Then, nonzero coupling constants for the interaction between electrons in the band 1 and the band 2 are,
\begin{eqnarray}
U^{aabb}_{\uparrow\downarrow\downarrow\uparrow},  \quad U^{aabb}_{\downarrow\uparrow\uparrow\downarrow},
\quad 
U^{abba}_{\uparrow\downarrow\downarrow\uparrow}, \quad U^{abba}_{\downarrow\uparrow\uparrow\downarrow},
\label{intap}
\end{eqnarray}
with $(a,b)=(1,2)$ or $(2,1)$.
For simplicity, we assume that all of these coupling constants are the same, and equal to $U_{12}$.
 Since we are concerned with effects of an applied magnetic field $H$, we also add the Zeeman term $-g\mu_{\rm B}S^zH$ to the Hamiltonian (\ref{eq:aham}).
Assuming the perfect nesting condition $\varepsilon_{\bm{k}+\bm{Q}_01}=-\varepsilon_{\bm{k}2}$, we write down
the non-interacting single-electron Green function for the band 1 with spin $\sigma$ in the SN state, $G_{\bm{k}1\sigma}(\varepsilon_n)$,
in the following forms,
\begin{eqnarray}
G_{\bm{k}1\sigma}(\varepsilon_n)=\frac{i\varepsilon_n+\varepsilon_{\bm{k}1}-s_{\sigma}h}{(i\varepsilon_n)^2-E^2_{k\sigma}},
\label{green1}
\end{eqnarray}
%\begin{eqnarray}
%G_{\bm{k}1\downarrow}(\varepsilon_n)
%=\frac{i\varepsilon_n+\varepsilon_{\bm{k}1}+h}{(i\varepsilon_n)^2-E^2_{k\uparrow}},
%\label{green2}
%\end{eqnarray}
where $h=g\mu_{\rm B}H$ and
 $E_{k\sigma}=\sqrt{(\varepsilon_{\bm{k}1}-s_{\sigma}h)^2+|\bm{d}_{12}(\bm{k})|^2}$ with 
 $s_{\sigma}=+1$ for $\sigma=\uparrow$ and $-1$ for $\sigma=\downarrow$.
 The Green functions for the band 2 are
 \begin{eqnarray}
&&G_{\bm{k}2\uparrow}(\varepsilon_n)=-G_{\bm{k}1\downarrow}(-\varepsilon_n), \label{eq:g2-1} \\
&&G_{\bm{k}2\downarrow}(\varepsilon_n)=-G_{\bm{k}1\uparrow}(-\varepsilon_n).
\label{eq:g2-2}
\end{eqnarray}
The non-interacting anomalous Green functions generated by the SN order are given by,
 \begin{eqnarray}
 F_{\bm{k}\uparrow\downarrow}(\varepsilon_n)=\int_0^{\beta} d\tau 
 \langle T\{c_{\bm{k}1\uparrow}c^{\dagger}_{\bm{k}2\downarrow}\}\rangle e^{i\omega_n\tau} 
 =\frac{-i|\bm{d}_{12}(\bm{k})|}{(i\varepsilon_n)^2-E^2_{k\uparrow}},
\label{angreen1}
 \end{eqnarray}
  \begin{eqnarray}
 F_{\bm{k}\downarrow\uparrow}(\varepsilon_n)=\int_0^{\beta} d\tau 
 \langle T\{c_{\bm{k}1\downarrow}c^{\dagger}_{\bm{k}2\uparrow}\}\rangle e^{i\omega_n\tau} 
=\frac{i|\bm{d}_{12}(\bm{k})|}{(i\varepsilon_n)^2-E^2_{k\downarrow}},
\label{angreen2}
 \end{eqnarray}

The longitudinal spin correlation function enhanced by nesting between the band $1$ and the band $2$ is
\begin{eqnarray}
\chi_{12}^{zz}(\bm{q},\omega_n)&=&\chi^{zz}_{1221}(\bm{q},\omega_n)+\chi^{zz}_{2112}(\bm{q},\omega_n)\nonumber \\
&+&
\chi^{zz}_{1212}(\bm{q},\omega_n)+\chi^{zz}_{2121}(\bm{q},\omega_n) \nonumber \\
&=&2[\chi^{zz}_{1221}(\bm{q},\omega_n)+\chi^{zz}_{2121}(\bm{q},\omega_n)],
\label{eq:spincor1}
\end{eqnarray}
where
\begin{eqnarray}
\chi_{abcd}^{zz}(\bm{q},\omega_n)=\int_0^{\beta} d\tau\langle T\{S^z_{ab}(\bm{q},\tau)S^z_{cd}(-\bm{q},0)\}\rangle e^{i\omega_n\tau},
\label{eq:spincor2}
\end{eqnarray}
with
 $S^z_{ab}(\bm{q})=\frac{1}{2}\sum_{\bm{k}}(c^{\dagger}_{\bm{k}+\bm{q}a\uparrow}c_{\bm{k}b\uparrow}-c^{\dagger}_{\bm{k}+\bm{q}a\downarrow}c_{\bm{k}b\downarrow})$.
In the second equality of (\ref{eq:spincor1}), we used the relations $\chi^{zz}_{1221}=\chi^{zz}_{2112}$ and
$\chi^{zz}_{1212}=\chi^{zz}_{2121}$.

Within the random phase approximation, $\chi_{12}^{zz}$ is given by
\begin{eqnarray}
\chi_{12}^{zz}(\bm{q},\omega_n)=\frac{\chi^{(0)}_{12}(\bm{q},\omega_n)}{1-U_{12}\chi^{(0)}_{12}(\bm{q},\omega_n)},
\label{eq:rpa-spin}
\end{eqnarray}
where %$U=U^{2211}_{\uparrow\downarrow\downarrow\uparrow}=U^{1122}_{\downarrow\uparrow\uparrow\downarrow}$,
%and 
$\chi^{(0)}_{12}(\bm{q},\omega_n)$ is the spin correlation function in the case without Coulomb interaction.
For $\bm{q}=\bm{Q}_0$, it is given by,
\begin{eqnarray}
&&\chi^{(0)}_{12}(\bm{Q}_0,\omega_n)=-\frac{T}{4}\sum_m\sum_{\bm{k}}[\sum_{\sigma=\uparrow,\downarrow}G_{\bm{k}1\sigma}(\varepsilon_m)
G_{\bm{k}2\sigma}(\varepsilon_m+\omega_n) \nonumber \\
&&-F_{\bm{k}\uparrow\downarrow}(\varepsilon_m)F_{\bm{k}\downarrow\uparrow}(\varepsilon_m+\omega_n)
-F_{\bm{k}\downarrow\uparrow}(\varepsilon_m)F_{\bm{k}\uparrow\downarrow}(\varepsilon_m+\omega_n) ] \nonumber \\
&&=\sum_{\bm{k}}\biggl[\frac{\tanh \frac{E_{\bm{k}\uparrow}}{2T}+\tanh \frac{E_{\bm{k}\downarrow}}{2T}}
{(E_{\bm{k}\uparrow}+E_{\bm{k}\downarrow})^2-(i\omega_n)^2}(E_{\bm{k}\uparrow}+E_{\bm{k}\downarrow})u_{+}(\bm{k})
 \nonumber \\
&&+\frac{\tanh \frac{E_{\bm{k}\uparrow}}{2T}-\tanh \frac{E_{\bm{k}\downarrow}}{2T}}
{(E_{\bm{k}\uparrow}-E_{\bm{k}\downarrow})^2-(i\omega_n)^2}(E_{\bm{k}\uparrow}-E_{\bm{k}\downarrow})u_{-}(\bm{k})\biggr],
\label{baresus}
\end{eqnarray}
where
\begin{eqnarray}
u_{\pm}(\bm{k})= \frac{1}{4}\left(1\pm\frac{\varepsilon_{\bm{k}1}^2+|\bm{d}_{12}(\bm{k})|^2-h^2}{E_{\bm{k}\uparrow}E_{\bm{k}\downarrow}}\right).
\end{eqnarray}
${\rm Im}\chi^{zz}(\bm{Q}_0,\omega)$ is approximately given by the retarded spin correlation function
${\rm Im}\chi_{12}^{zz R}(\bm{Q}_0,\omega)$.

In a similar manner, we can obtain $\chi^{zz}(\bm{Q_1},\omega)$, which is dominated by the longitudinal spin correlation function enhanced by nesting between 
the band 2 and the band 3, 
\begin{eqnarray}
\chi_{23}^{zz}(\bm{q},\omega_n)=2[\chi^{zz}_{2332}(\bm{q},\omega_n)+\chi^{zz}_{2323}(\bm{q},\omega_n)].
\label{chi23}
\end{eqnarray}
From Eqs.(\ref{orb2}) and (\ref{orb3}) and the conservation of $J_z$ in interaction processes, 
we see that nonzero coupling constants of the interaction between electrons in the band 2 and the band 3
are given by Eq.(\ref{intap}) with $(a,b)=(2,3)$ or $(3,2)$.  We assume that all of them are equal to $U_{23}$.
To simplify the calculation, 
we also impose the perfect nesting condition $\varepsilon_{\bm{k}+\bm{Q}_12}=-\varepsilon_{\bm{k}3}$.
Since the energy gap does not open in the band 3 in our model, we obtain
within the random phase approximation,
\begin{eqnarray}
\chi_{23}^{zz}(\bm{q},\omega_n)=\frac{\chi^{(0)}_{23}(\bm{q},\omega_n)}{1-U_{23}\chi^{(0)}_{23}(\bm{q},\omega_n)},
\label{eq:rpa-spin23}
\end{eqnarray}
where
\begin{eqnarray}
\chi^{(0)}_{23}(\bm{q},\omega_n)=\sum_{\bm{k},\sigma}\biggl[\frac{\tanh\frac{E_{\bm{k}\sigma}}{2T}
+\tanh\frac{\varepsilon_{\bm{k}1}+s_{\sigma}h}{2T}}{i\omega_n+E_{\bm{k}\sigma}+\varepsilon_{\bm{k}1}+s_{\sigma}h}
w_{+}(\bm{k}) \nonumber \\
-\frac{\tanh\frac{E_{\bm{k}\sigma}}{2T}
-\tanh\frac{\varepsilon_{\bm{k}1}+s_{\sigma}h}{2T}}{i\omega_n-E_{\bm{k}\sigma}+\varepsilon_{\bm{k}1}+s_{\sigma}h}
w_{-}(\bm{k})
\biggr],
\label{baresus2}
\end{eqnarray}
with
\begin{eqnarray}
w_{\pm}(\bm{k})=\frac{1}{4}\left(1\pm\frac{\varepsilon_{\bm{k}1}-s_{\sigma}h}{E_{\bm{k}\sigma}}\right).
\end{eqnarray}
${\rm Im}\chi^{zz}(\bm{Q}_1,\omega)$ is approximated by ${\rm Im}\chi_{23}^{zz R}(\bm{Q}_1,\omega)$.

Plots of ${\rm Im}\chi_{12}^{zz R}(\bm{Q}_0,\omega)$ and
${\rm Im}\chi_{23}^{zz R}(\bm{Q}_1,\omega)$ as functions $\omega$ are shown in FIG.1(b) and (c), respectively, in the main text.

\subsection{B. Microscopic mechanism of the spin nematic state}

In this section, we discuss microscopic mechanism of the SN order for URu$_2$Si$_2$.
We consider a scenario that spin fluctuations and orbital fluctuations associated with the Fermi surface nesting with the nesting vectors
$\bm{Q}_0$ and $\bm{Q}_1$ 
give rise to the phase transition to the SN order.
Our argument is based on the three band model (\ref{eq:aham}). 
%Generally, $U^{abcd}_{\sigma_1\sigma_2\sigma_3\sigma_4}$ depends on momenta through the form factors raised by the basis transformation.
%However, we neglected them. This does not affect the following qualitative argument.

The spin fluctuations and the orbital fluctuations with the propagating wave numbers 
$\bm{Q}_{0}$ and $\bm{Q}_{1}$ mediate the effective interactions 
between electrons, 
\begin{eqnarray}
\mathcal{S}_{\rm int}=\sum_{k,k',q\atop a,b,c,d,\sigma_i}\Gamma^{abcd}_{\sigma_1\sigma_2\sigma_3\sigma_4}(q)c^{\dagger}_{k+q a\sigma_1}c^{\dagger}_{k'-q b\sigma_2}
c_{k' c\sigma_3}c_{k d\sigma_4},
\end{eqnarray} 
with $q=(\bm{q},\omega)$. 

From Eqs.(\ref{orb1}), (\ref{orb2}), and (\ref{orb3}), and the conservation of 
the total $J_z$ in each scattering process, 
the effective interaction that mediates the particle-hole pairing characterizing the SN order is 
approximately given by,
\begin{eqnarray}
\Gamma^{2211}_{\uparrow\downarrow\downarrow\uparrow}(q)&=&
\Gamma^{2211(0)}_{\uparrow\downarrow\downarrow\uparrow}(q)
-g^{1212}_{\uparrow\uparrow\uparrow\uparrow}g^{2211}_{\downarrow\uparrow\uparrow\downarrow} 
[\chi^{12}_{\uparrow\uparrow}(q)+\chi^{12}_{\downarrow\downarrow}(q)] \nonumber \\
&&-g^{2231}_{\uparrow\uparrow\uparrow\uparrow}g^{2321}_{\downarrow\uparrow\uparrow\downarrow}
[\chi^{23}_{\uparrow\uparrow}(q)+\chi^{23}_{\downarrow\downarrow}(q)].
\label{eq:eff-int}
\end{eqnarray}
Here, the second and third terms are the interactions mediated via the orbital fluctuations which is characterized by
the orbital susceptibility,
\begin{eqnarray}
\chi^{ab}_{\sigma\sigma}(q)=\int^{\beta}_0 d\tau \langle T\{N_{ab\sigma}(\bm{q},\tau)N_{ba\sigma}(-\bm{q},0)\}\rangle e^{i\omega_n\tau},
\label{orbsus}
\end{eqnarray}
where $N_{ab\sigma}(\bm{q})=\sum_{\bm{k}}c^{\dagger}_{\bm{k}+\bm{q}a\sigma}c_{\bm{k}b\sigma}$.
$g^{1212}_{\uparrow\uparrow\uparrow\uparrow}$, $g^{2211}_{\downarrow\uparrow\uparrow\downarrow}$, 
$g^{2321}_{\uparrow\downarrow\downarrow\uparrow}$ and $g^{2321}_{\downarrow\uparrow\uparrow\downarrow}$
are effective coupling constants for interactions between the orbital fluctuations and electrons, which preserve  $J_z$.
The first term of (\ref{eq:eff-int}) is the other interaction processes.
The Fermi surface
nesting between the band 1(3) and the band 2 with the nesting vector $\bm{Q}_0$ ($\bm{Q}_1$) 
 enhances the orbtial-fluctuation-mediated part of (\ref{eq:eff-int}), while the first term of
(\ref{eq:eff-int}) is not much affected by it.
In Eq.(\ref{eq:eff-int}), we have neglected the interaction processes that are mediated by orbital fluctuations associated with
$\chi^{ab}_{\uparrow\downarrow}(q)$ and $\chi^{ab}_{\downarrow\uparrow}$, because these orbital fluctuations
accompany transverse spin fluctuations, and, according to neutron scattering measurements,
transverse spin fluctuations are fully suppressed in URu$_2$Si$_2$.\cite{PhysRevB.43.12809}
The second and third terms of (\ref{eq:eff-int}) give rise to the particle-hole pairing interaction in the $d_{\mu'z}$-wave channel
$V_d\equiv 
\sum_{\bm{k},\bm{k}'}^{\rm F.S.}\phi(\bm{k})\phi(\bm{k}')\Gamma^{2211}_{\uparrow\downarrow\downarrow\uparrow}
(\omega_n,\bm{k}-\bm{k}')$. 
Here $\sum_{k,k'}^{\rm F.S.}$ denotes the momentum sum over the Fermi surfaces.
Since $\sum_k\phi(\bm{k})\phi(\bm{k}+\bm{Q}_m)<0$ with $m=0,1$, 
and the orbital susceptibility (\ref{orbsus}) is significantly enhanced
at $\bm{q}=\bm{Q}_1$ (or $\bm{Q}_0$), the sign of $V_d$ is positive, which leads to the $d$-wave particle-hole pairing characterizing the SN state.

In addition to (\ref{eq:eff-int}), there is another important interaction channel which is related to the instability toward
antiferromagnetism with staggered magnetization parallel to the $z$-axis. This instability is also  
driven by the Fermi surface nesting
between the band 1 and the band 2, as in the case of the SN order \cite{oppen}.
We denote this effective interaction as $\Gamma^{2211}_{\uparrow\downarrow\uparrow\downarrow}(q)$.
Note that $\Gamma^{2211}_{\uparrow\downarrow\uparrow\downarrow}(q)$ 
is not enhanced by the orbital fluctuations associated with the orbital correlation function (\ref{orbsus}).
Thus, if the orbital fluctuations are sufficiently strong, the effective interaction (\ref{eq:eff-int}) gives rise to the instability toward
the SN state.
The transition temperature of the SN order
within a mean field approximation is determined by the self-consistent gap equation,
\begin{eqnarray}
\Delta_1=V_d\sum_{\bm{k}}\phi^2(\bm{k})\frac{\tanh(\frac{E_{{\bm k}}}{2T})}{E_{{\bm k}}}\Delta_{1} \biggl/\sum_{\bm{k}}\phi^2(\bm{k}).
\label{gap-eq}
\end{eqnarray}

When there is an applied magnetic field parallel to the $z$-axis, the orbital susceptibility (\ref{orbsus}) is reduced.
For a small field $H_z$, the nonzero lowest order correction is $\delta\chi^{ab}_{\sigma\sigma}(q)\sim -c_0(q)H^2_z $
with $c_0(q)$ a function of $q$.
The magnetic field parallel to the $z$-axis affects the self-consistent gap equation (\ref{gap-eq})
only through the effective pairing interaction.
Thus, the transition temperature in the case with $H_z$ behaves like
$T_{\rm HO}(H_z)\sim T_{\rm HO}(0)-a_0 H_z^2$ with $a_0$ a positive constant.
This field dependence of the transition temperature is actually observed in experiments \cite{dijk}. 
 
\subsection{C. Specific heat jump}

In this section, we discuss the specific heat jump at the HO transition point.
According to specific heat measurements under magnetic fields applied along the $z$-axis,
the specific heat jump $\Delta C$ at $T=T_{\rm HO}$ is almost constant as a function of the magnetic fields,
though the transition temperature decreases as $T_{\rm HO}(H_z)\sim T_{\rm HO}(0)-a_0H_z^2$, as mentioned in the previous section \cite{dijk,jaime}.  
The constant specific heat jump is unusual for the itinerant f-electron system, since in the Fermi liquid state, the entropy generally decreases as $T$ decreases.
However, here, we show that this remarkable feature is explained by taking into account 
the selfenergy corrections arising from AF spin fluctuations.
We start from the Ginzburg-Landau (GL) free energy in the vicinity of the HO transition point,
$\mathcal{F}=a\psi^2+b\psi^4
$, with $\psi$ the order parameter for the SN state. 
We have chosen the spin quantization axis parallel to $\bm{d}_{12}$. Thus, $\psi=\Delta_1$ is a scalar order parameter.
The specific heat jump at $T=T_{\rm HO}$ is given by
\begin{eqnarray}
\Delta C=\frac{T_{\rm HO}}{2b}\left( \frac{\partial a}{\partial T}\right)^2\biggl|_{T=T_{\rm HO}}.
\label{eq:sphj}
\end{eqnarray}
The coefficients of the GL free energy $a$ and $b$ are calculated from the normal Green functions,
\begin{eqnarray}
G_{\bm{k}1\sigma}(\varepsilon_n)=\frac{1}{Z_{\bm{k}}(\varepsilon_n)i\varepsilon_n-\varepsilon_{\bm{k}1}}
\end{eqnarray}
with 
\begin{eqnarray}
Z_{\bm{k}}(\varepsilon_n)=1-\frac{1}{i\varepsilon_n}\Sigma_{\bm{k}}(\varepsilon_n),
\end{eqnarray}
and $\Sigma_{\bm{k}}(\varepsilon_n)$ the selfenergy, 
and $G_{\bm{k}2\sigma}(\varepsilon_n)$ which is obtained from the relations (\ref{eq:g2-1}) and (\ref{eq:g2-2}).
The results are
\begin{eqnarray}
a=\frac{1}{g}-\Pi(\bm{Q}_0),
\end{eqnarray}
with 
\begin{eqnarray}
\Pi(\bm{Q}_0)=-T\sum_{|n|\leq n_c}\sum_{\bm{k}}\phi^2(\bm{k})G_{\bm{k}1\sigma}(\varepsilon_n)G_{\bm{k}+\bm{Q}_02-\sigma}(\varepsilon_n),
\label{eq:glca}
\end{eqnarray}
and
\begin{eqnarray}
b=T\sum_{|n|\leq n_c}\sum_{\bm{k}}\phi^4(\bm{k})
[G_{\bm{k}1\sigma}(\varepsilon_n)G_{\bm{k}+\bm{Q}_02-\sigma}(\varepsilon_n)]^2.
\label{eq:glcb}
\end{eqnarray}
Here $n_c=E_c/T$ with $E_c$ an energy cutoff for the effective particle-hole pairing interaction $g$.
For simplicity, we neglect $\bm{k}$-dependence of $Z_{\bm{k}}(\varepsilon_n)$.
Then, from eqs.(\ref{eq:sphj}), (\ref{eq:glca}), and (\ref{eq:glcb}),
we obtain,
\begin{eqnarray}
\Delta C \approx N(0)Z(E_c)T_{\rm HO},
\label{eq:delc}
\end{eqnarray}
where $N(0)$ is the density of states at the Fermi level.
The selfenergy corrections $Z(E_c)$ mainly stem from interactions with AF spin fluctuations and orbital fluctuations.
Note that the AF spin fluctuations expressed by (\ref{eq:spincor1}) and (\ref{chi23}) and
the orbital fluctuations characterized by (\ref{orbsus}) are not independent, but deeply correlated.
The former contributions generally enhance the latter, and vice versa.
We postulate that the normal selfenergy corrections are mainly due to the interaction with the AF spin fluctuations
with $\bm{Q}_0$, while the particle-hole pairing interaction (\ref{eq:eff-int}) is dominated by
the fluctuations with $\bm{q}\sim \bm{Q}_1$; i.e. in Eq.(\ref{eq:eff-int}), the coupling strength of the second term is 
much smaller than that of the third one, though the correlation length for the spin fluctuations with
$\bm{Q}_0$ is larger than that with $\bm{Q}_1$.
We also assume that  the characteristic energy scale of the AF spin fluctuations with $\bm{q}=\bm{Q}_0$
is sufficiently larger than the Zeeman energy of the applied fields, and
the AF spin fluctuations with $\bm{Q}_0$ is not much affected by the magnetic fields, at least, 
in the vicinity of $T\sim T_{\rm HO}$. 
The most important effect on $Z(\varepsilon_n)$ in the vicinity of the transition temperature
is the quasiparticle damping $\gamma$.
If the system is sufficiently clean, the quasiparticle damping raised by the interaction with the three-dimensional AF spin fluctuations with $\bm{Q}_0$ is $\gamma \sim \xi^2[\varepsilon_n^2+(\pi T)^2]$ with $\xi$ the correlation length for the AF spin fluctuations \cite{tada}. Then, from (\ref{eq:delc}), we have 
$\Delta C\sim \xi^2T_{\rm HO}$ up to a constant factor.
If the system is in the vicinity of the AF quantum critical point, i.e.
$\xi^2 \sim 1/T$, $\Delta C$ does not depend on $T_{\rm HO}$.
This result provides a possible explanation for the above-mentioned experimental observation that
the specific heat jump at $T=T_{\rm HO}$ is not changed by the magnetic field,
though the transition temperature is decreased by the magnetic field as
$T_{\rm HO}(H_z)\sim T_{\rm HO}(0)-a_0H_z^2$
\cite{dijk,jaime}.
We note that the above scenario of the constant specific heat jump is valid as long as $T_{\rm HO}(H_z)$ is sufficiently large compared to the change of the characteristic energy of the spin fluctuation raised by the magnetic field.
Thus, when $T_{\rm HO}(H_z)$ is decreased too much by the applied field, the deviation from the above result may appear.

\end{document}